\documentclass[a4paper,12pt]{article}
\usepackage[cp1251]{inputenc}
\usepackage[english]{babel}
\usepackage[dvips]{graphicx}
\graphicspath{{.}}

\begin{document}
\mathsurround=2pt \sloppy
\title{\bf  The distorted axi-planar superfluid phase of $^3$He in the "nematically ordered" aerogel}
\author{I. A. Fomin$^1$  and E.V. Surovtsev$^{1,2}$
\vspace{.5cm}\\
{\it $^1$P. L. Kapitza Institute for Physical Problems}\\ {\it Russian Academy of Science},\\{\it
Kosygina 2,
 119334 Moscow, Russia}\\  $^{2}${\it Moscow Institute of Physics and
Technology}, \\{\it Dolgoprudny, Mosow region}}
\maketitle
\begin{abstract}
Stimulated by the results of NMR experiments with superfluid $^3$He in "nematically ordered" aerogel \cite{askh} we report on the results of phenomenological analysis of stability of different  phases of superfluid $^3$He subjected to a strong homogeneous   uniaxial anisotropy. On a basis of this analysis we suggest a form of the order parameter for the new ESP2 phase observed in the quoted experiments. In the weak coupling limit the suggested order parameter approaches that of the axi-planar phase. We discuss a possible experimental check of the proposed identification of the new phase.
\end{abstract}

\textbf{1.} Aerogels opened a possibility to impose a global orbital anisotropy on the p-wave Fermi superfluid -- $^3$He  \cite{HS}. In  the mean field approximation deformed aerogels couple to the orbital part of the order parameter in a way analogous to Zeeman coupling of the spin part to magnetic field. Corresponding term in the density of free energy is proportional to $\kappa_{jl}A_{\mu j}A_{\mu l}^*$. The complex 3$\times$3 matrix $A_{\mu j}$ is the order parameter of $^3$He, averaged over distances larger than the distance between the strands of aerogel $\xi_a$. The index $\mu$ stands for spin projections and $j$ -- for orbital. Real symmetric traceless tensor $\kappa_{jl}$ describes  effect of the anisotropy. If anisotropy is uniaxial and $\hat{z}$ is the symmetry axis $\kappa_{jl}$ can be written as  $\kappa_{xx}=\kappa_{yy}\equiv -\kappa$, $\kappa_{zz}=2\kappa$. The uniaxial anisotropy splits the transition temperature for orbital triplet ($l$=1) in two different temperatures-- for $l_z=0$ and for $l_z=\pm 1$, favoring one or the other state, depending on the sign of $\kappa$. Uniaxial compression  ($\kappa>0$) favors $l_z=\pm 1$ projections. It increases stability of the ABM phase and orients its order parameter. Physical manifestations of this orientation were discussed in the literature \cite{kunim,dm}. Uniaxial stretch ($\kappa<0$) favors $l_z=0$ projection, which taken alone can not form the orbital part of the ABM order parameter. Aoyama and Ikeda \cite{AI} predicted that this type of anisotropy will create just below $T_c$ a region of stability of the polar phase. The width of this region depends on the absolute value of $\kappa$, i.e. on a rate of anisotropy of aerogel.  Possibility of stretching of silica aerogels is limited. To obtain an appreciable anisotropy in the experiments \cite{askh} a new type of aerogel was used \cite{askh1},\cite{askh2}. Its strands are nearly parallel to one direction (anisotropy axis), so that it was dubbed as "nematically ordered". Anticipating we remark that even for that new aerogel the estimated from the experiment value of $\kappa$ is of the order of 10$^{-2}$ and the anisotropy has strong effect on the phase diagram of  $^3$He in aerogel only near to its transition temperature $T_{ca}$.

The authors of the paper \cite{askh} investigated superfluid phases of $^3$He in aerogel by CW and pulsed NMR methods for different orientations of the DC magnetic field with respect to the anisotropy axis (taken as $z$-axis in what follows). Several phases with different NMR-signatures were observed and the phase diagram of $^3$He in "nematically ordered" aerogel in the temperature-pressure coordinates was constructed. The results, obtained on cooling from the normal phase confirm the scenario, suggested by Aoyama and Ikeda.  Helium first enters the Equal Spin Pairing (ESP) state which can be interpreted as following the root from the polar phase via the second order phase transition to the ABM-phase with polar distortion. The latter on further cooling jumps to the low-temperature phase (LTP). This phase has the  NMR properties of the BW phase with the polar distortion.

Surprising results are obtained on warming from the LTP. On its approach to the $T_c$ helium jumps to the ESP-phase which has different NMR frequency shift from one observed on cooling. The new phase was denoted as ESP2 and it is not properly identified. To clarify this situation we carried out an analysis of possible phase transitions in uniaxially stretched aerogel phenomenologically, within the Landau theory of phase transitions. Superfluid $^3$He in aerogel can be considered as a uniform media if the Ginzburg and Landau coherence length $\xi(T)$ is greater than the average distance $\xi_a$ between the strands of aerogel. In terms of the zero temperature coherence length $\xi_0$ this condition can be reformulated as a restriction of the temperature interval where the  state of helium can be characterized by average order parameter: $(T-T_{ca})/T_{ca}\ll (\xi_0/\xi_a)^2$. For the low pressure data of Ref.\cite{askh} (P$<$6.5 bar) the Landau theory can be applied down to $\approx .85 T_{ca}$. This region includes the most interesting phase transitions. For the high pressure data (P$>$12 bar) $\xi_0\ll\xi_a$ and the interval of applicability of uniform approximation is several times smaller. Outside of this interval it can be used only as a qualitative guidance because the state becomes essentially nonuniform and local fluctuations of the order parameter can render significant contribution.

\textbf{2.}  Within  the mean field approximation the average thermodynamic potential of the p-wave superfluid in a globally anisotropic  aerogel has the following form:
$$
 \Phi_s=\Phi_n+N_{eff}[(\tau\delta_{jl}+\kappa_{jl}) A_{\mu j}A_{\mu j}^*+
\frac{1}{2}(\beta_1A_{\mu j}A_{\mu j}A_{\nu l}^*A_{\nu l}^*+\beta_2A_{\mu j}A_{\mu j}^*A_{\nu l}A_{\nu l}^*+
$$
$$\beta_3A_{\mu j}A_{\nu j}A_{\mu l}^*A_{\nu l}^*+\beta_4A_{\mu j}A_{\nu j}^*A_{\nu l}A_{\mu l}^*+
\beta_5A_{\mu j}A_{\nu j}^*A_{\mu l}A_{\nu l}^*)]       \eqno(1)
$$
Here $\tau=(T-T_{c0})/T_{c0}$ is the dimensionless temperature counted from the transition temperature at $\kappa=0$, $N_{eff}$ denotes an "effective density of states".
The values of the phenomenological coefficients  $\beta_1,...\beta_5$  depend on pressure and properties of aerogel.   The values of $\beta_1,...\beta_5$ derived from the BCS theory are referred as the weak coupling limit. When normalized to $\beta_2$ these values are  $\beta_1,\beta_2,\beta_3,\beta_4,\beta_5$=$\beta_2$(-1/2,1,1,1,-1) \cite{VW}.
Experimental data indicate that real values of these ratios even for the bulk $^3$He, may deviate from their weak coupling values for 20 \% -- 30 \% \cite{halp2}. The deviations are important at determining regions of stability of different phases.

For uniaxially anisotropic aerogel the second order terms in Eq. (1) have the form
$(\tau+2\kappa)A_{\mu z}A_{\mu z}^*+(\tau-\kappa)(A_{\mu x}A_{\mu x}^*+A_{\mu y}A_{\mu y}^*)$. If $\kappa<0$ the superfluid  transition takes place at $\tau=-2\kappa$. Below this $\tau$ the order parameter is a complex spin vector $A_{\mu z}=\Delta_0A^{(0)}_{\mu z}$ with  $A^{(0)}_{\mu z}$  normalized by the condition $A^{(0)}_{\mu z}A^{(0)\ast}_{\mu z}=1$. Minimization of the free energy with that   order parameter renders
$$
\Delta_0^2=-\frac {\tau+2\kappa}{\beta_{15}|A^{(0)}_{\mu z}A^{(0)}_{\mu z}|^2+\beta_{234}}.    \eqno(2)
$$
Here the conventional shorthand notations $\beta_{15}=\beta_1+\beta_5$ etc. are used. All data indicate that in $^3$He $\beta_{15}<0 $ , then the maximum gain of energy is reached at $|A^{(0)}_{\mu z}A^{(0)}_{\mu z}|^2=1$. Together with the normalization condition it renders $A^{(0)}_{\mu z}=\exp(i\varphi)d_{\mu}$, where  $d_{\mu}$ is a real spin vector. With the account of the orbital part it reproduces the order parameter of the polar phase $A_{\mu j}^{0}=\Delta_0\exp(i\varphi)d_{\mu}m_j$, where $m_j$ is a unit vector in $z$-direction in agreement with the Ref.\cite{AI}. For the overall  amplitude we have from Eq. (2): $\Delta_0^2=-(\tau+2\kappa)/\beta_{12345}$.

\textbf{3.}  To find further possible phase transitions we represent the order parameter as $A_{\mu j}=A_{\mu j}^{0}+a_{\mu j}$, where $a_{\mu j}$ is a small increment, and expand the change of the thermodynamic potential
$\bar{\Phi}\equiv \Phi_s-\Phi_n$ in powers of $a_{\mu j}$. All experimentally observed transitions take place at $|\tau|<.15$, where Eq.(1) is still a good approximation for $\bar{\Phi}$. Temperature of the transition is determined by the second order terms:
$$
\bar{\Phi}(A_{\mu j},A_{\mu j}^*)=\bar{\Phi}(A_{\mu j}^0,A_{\mu j}^{0*})+
$$
$$
\frac{1}{2}\left\{\frac{\partial^2\bar{\Phi}}{\partial A_{\mu j}\partial A_{\nu l}}a_{\mu j}a_{\nu l}+
2\frac{\partial^2\bar{\Phi}}{\partial A_{\mu j}\partial A^*_{\nu l}}a_{\mu j}a_{\nu l}^*+\frac{\partial^2\bar{\Phi}}{\partial A^*_{\mu j}\partial A^*_{\nu l}}a^*_{\mu j}a_{\nu l}^*\right\}.                          \eqno(3)
$$
At the transition the linear equation
$$
\frac{\partial^2\bar{\Phi}}{\partial A_{\mu j}\partial A_{\nu l}}a_{\nu l}+\frac{\partial^2\bar{\Phi}}{\partial A_{\mu j}\partial A^*_{\nu l}}a_{\nu l}^*=0                                                 \eqno(4)
$$
together with its complex conjugated equation acquires a nontrivial solution, which breaks the symmetry of the polar phase.
To select essential solutions we have to impose the orthogonality condition $A_{\mu j}^{0*}a_{\mu j}=0$, or explicitly: $d_{\mu}m_ja_{\mu j}=0$. Because of the degeneracy of $A_{\mu j}^{0}$ with respect to orientation of the spin part $d_{\mu}$ even more restrictive  condition can be imposed $m_ja_{\mu j}=0$. Using the freedom in determining the overall gauge we require that for the polar phase
$\exp(i\varphi)=1$, then $A_{\mu j}^{0}$ is real. Eq. (4) can be rewritten as two equations separately for the real and imaginary parts of $a_{\mu j}=a^R_{\mu j}+ia^I_{\mu j}$. Both equations have the form $K_{\mu\nu j l}^{R,I}a^{R,I}_{\mu j}=0$ with
$K_{\mu\nu j l}^{R,I}=A^{R,I}\delta_{\mu\nu}\delta_{jl}+B^{R,I}\delta_{\mu\nu}m_jm_l+C^{R,I}d_{\mu}d_{\nu}\delta_{jl}+D^{R,I}d_{\mu}d_{\nu} m_jm_l$. Because of the imposed conditions the terms with $B^{R,I}$ and $D^{R,I}$ do not contribute to the equations for $a^{R,I}_{\mu j}$. The resulting equations have the form:
$$
(A^{R,I}\delta_{\mu\nu}+C^{R,I}d_{\mu}d_{\nu})a^{R,I}_{\nu j}=0.                    \eqno(5)
$$
Let $d_{\mu},e_{\mu},f_{\mu}$ be orthogonal basis in spin space. Taking projections of Eq. (5) on each of these vectors we arrive  at the following possibilities: 1) $A^{R,I}+C^{R,I}=0$,  $d_{\mu}a^{R,I}_{\mu j}\neq 0$ or 2) $A^{R,I}=0$,  $e_{\mu}a^{R,I}_{\mu j}\neq 0$, and  $f_{\mu}a^{R,I}_{\mu j}\neq 0$. In both cases the orbital part is orthogonal to $m_j$. Coefficients $A^{R,I}$ and $C^{R,I}$ can be expressed in terms of $\tau, \kappa$ and  $\beta_1,...\beta_5$ with the aid of Eqns. (1),(2).

First consider the imaginary part. The sum  $A^I+C^I=\tau-\kappa+(\beta_{245}-\beta_{12})\Delta_0^2$ turns to zero at $\tau=\tau_A$, $\tau_A=\kappa(3\beta_{245}-\beta_{13})/2\beta_{13}$. The emerging component is
$$
a^I_{\mu j}=i\Delta_1d_{\mu}n_j\prime                                    \eqno(6)
$$
where $n_j\prime$ is a unit vector, orthogonal to $m_j$ and $\Delta_1$ is a real amplitude.
Together with  $A_{\mu j}^{0}$ it forms the order parameter of the ABM-phase with a polar distortion  $A_{\mu j}=d_{\mu}(\Delta_0m_j+\Delta_1n_j\prime)$, which was discussed in the Refs. \cite{AI},\cite{askh}.
The temperature dependence of $\Delta_0$ and $\Delta_1$ at $\tau<\tau_A$ is found by minimization of $\Phi_s$ (Eq. (1)):
$\Delta_1^2=-\frac{\tau-\tau_A}{2\beta_{245}}$, $\Delta_0^2=-\frac{3\kappa}{2\beta_{13}}-\frac{\tau-\tau_A}{2\beta_{245}}$.
In comparison with the polar phase the new phase has lower thermodynamic potential. The gain is
$\Phi_A-\Phi_p=-\frac{\beta_{13}}{2\beta_{245}\beta_{12345}}(\tau-\tau_A)^2$.  The condition 2) for imaginary part $A^I=0$ is not satisfied at small $\tau$.

For the real part the sum  $A^R+C^R=-3\kappa$ is finite at all $\tau$. It means that the real part of the projection of $a_{\mu j}$ on $d_{\mu}$ is absent. On the contrary $A^R=\tau-\kappa+\beta_{12}\Delta_0^2$ turns to zero at $\tau=\tau_B$ , $\tau_B=\kappa(1+3\beta_{12}/\beta_{345})$. Below this temperature finite projection of $a^R_{\mu j}$ on $e_{\mu}$ and $f_{\mu}$ can exist. With the aid of rotations around $d_{\mu}$ and  $m_j$ the 2$\times$2 real tensor in a space orthogonal to these vectors can be transformed to the form
$$
a^R_{\mu j}=\Delta_2e_{\mu}l_j+\Delta_3f_{\mu}n_j,                                    \eqno(7)
$$
$n_j$ and $l_j$ are two mutually orthogonal vectors, forming together with $m_j$  basis in the orbital space, $\Delta_2$ and $\Delta_3$ are real amplitudes.
Minimization of $\Phi_s$ at  $\tau<\tau_B$ renders:  $\Delta_2=\Delta_3$, then  $A_{\mu j}^{0}+a^R_{\mu j}$ forms the order parameter of the BW-phase with a polar distortion, also discussed in Refs. \cite{AI},\cite{askh}. Temperature dependence of the amplitudes is given by $\Delta_2^2=-\frac{\tau-\tau_B}{3\beta_{12}+\beta_{345}}$, $\Delta_0^2=-\frac{3\kappa}{\beta_{345}}+\Delta_2^2$ with the energy gain $\Phi_B-\Phi_p=-\frac{\beta_{345}}{\beta_{12345}(3\beta_{12}+\beta_{345})}(\tau-\tau_B)^2$.

On cooling from the polar phase the perturbation with the higher transition temperature $\tau$ occurs first. According to the experiment \cite{askh} this is the ESP-phase. Of the two considered possibilities only $a^I_{\mu j}$ meets this requirement.  It means that $\tau_A>\tau_B$ and the ABM-phase with a polar distortion develops below $\tau_A$.
Repetition of the same argument with $A_{\mu j}^{(0)}=\Delta_0d_{\mu}m_j+i\Delta_1d_{\mu}n_j$ shows that no further continuous transition occurs on cooling within the limits of applicability of the expansion Eq. (1). The more advantageous at low temperatures B-phase with a polar distortion occurs via the first order transition in agreement with the experiment \cite{askh} and with the theoretical analysis of Ref. \cite{AI}. Since the sequence of the transitions depends on the sign of the difference
$\tau_A-\tau_B=3\kappa\beta_{12345}(1/2\beta_{13}-1/\beta_{345})$ we conclude that for $^3$He in nematically ordered aerogel $\beta_{345}<2\beta_{13}$.

\textbf{4.} The case $\tau_A=\tau_B\equiv\tau_{AB}$ is special. It is realized in particular for the weak coupling values of the coefficients $\beta_1,...\beta_5$. In this case the emerging increment of the order parameter can be searched as a linear combination of $a^I_{\mu j}$ and $a^R_{\mu j}$ so that the order parameter just below the transition  has a form:
$$
A_{\mu j}=\Delta_0d_{\mu}m_j+i\Delta_1d_{\mu}(n_j\sin\theta+l_j\cos\theta)+\Delta_2e_{\mu}l_j+\Delta_3f_{\mu}n_j,  \eqno(8)
$$
where $\theta$ is the angle between $n\prime_j$ and $l_j$.
Minimization of thermodynamic potential (1) with the weak coupling values of $\beta_1,...\beta_5$ renders both states considered above plus one more ESP-state:
$$
A_{\mu j}^{AP}=\Delta_0d_{\mu}m_j+i\Delta_1d_{\mu}n_j+\Delta_2e_{\mu}l_j,  \eqno(9)
$$
which is a particular form of the axi-planar phase \cite{merm},\cite{gould}. For the amplitudes  $\Delta_0,\Delta_1,\Delta_2$  we  have: $\Delta_1^2=\frac{1}{8\beta_2}\left(\frac{5\kappa}{2}-\tau\right)$, $\Delta_2^2=3\Delta_1^2$, $\Delta_0^2=-\frac{1}{4\beta_2}(2\tau+7\kappa)$ and the energy gain $(\Phi_p-\Phi_{AP})/N_{eff}=\frac{1}{6\beta_2}(\tau-\tau_A)^2$ - i.e. exactly the same as that for the distorted ABM-phase. At $\tau_A=\tau_B$ both ESP phases are less advantageous than the distorted B-phase, so that the B-like distortion develops at $\tau<\tau_{AB}$.

Deviation of the coefficients $\beta_s$ from their weak coupling values results in splitting of the transition. If the splitting is small i.e. $(\tau_A-\tau_B)\ll\tau_A$, the axi-planar order parameter Eq. (9) with possibly corrections of the order of $(\tau_A-\tau_B)/\tau_A$ in a region $|\tau-\tau_A|\gg (\tau_A-\tau_B)$ renders an asymptotic form for a local minimum of the thermodynamic potential, which can be reached via the first order transition or a sequence of transitions. To find out  properties of the new phase we have to substitute the order parameter (8) in the  expression for the thermodynamic potential Eq. (1) and minimize it over 5 free parameters: $\theta, \Delta_0, \Delta_1, \Delta_2, \Delta_3$. Extremum over $\theta$ is reached at: 1) $\Delta_1^2=0$, 2) $\Delta_2^2=\Delta_3^2$,  3) $\sin\theta=0$ or $\cos\theta=0$ (these two possibilities are equivalent up to notations). The possibilities 1), 2) correspond to the distorted BW-phase.
The most interesting for the present discussion is the third possibility. For definiteness we take $\cos\theta=0$. If all $\Delta_s^2$ are finite they have to obey the following system of linear equations:
$$
\beta_{12345}\Delta_0^2+(\beta_{245}-\beta_{13})\Delta_1^2+\beta_{12}\Delta_2^2+\beta_{12}\Delta_3^2=-(\tau+2\kappa),  \eqno(10) $$
$$
(\beta_{245}-\beta_{13})\Delta_0^2+\beta_{12345}\Delta_1^2+(\beta_2-\beta_1)\Delta_2^2+(\beta_{234}-\beta_{15})\Delta_3^2=
-(\tau-\kappa),      \eqno(11)
$$
$$
\beta_{12}\Delta_0^2+(\beta_2-\beta_1)\Delta_1^2+\beta_{12345}\Delta_2^2+\beta_{12}\Delta_3^2=-(\tau-\kappa),    \eqno(12)
$$
$$
\beta_{12}\Delta_0^2+(\beta_{234}-\beta_{15})\Delta_1^2+\beta_{12}\Delta_2^2+\beta_{12345}\Delta_3^2=-(\tau-\kappa).   \eqno(13)
$$
The lack of precise knowledge of the values of $\beta$-coefficients introduces ambiguity in analysis of these equations.  Here we remark only on some properties, which follow from the assumption that the strong coupling corrections are small. Combination of Eqns. (12) and (13) renders the relation:
$$
(\beta_{34}-\beta_{5})\Delta_1^2=\beta_{345}(\Delta_2^2-\Delta_3^2).        \eqno(14)
$$
As a result if $\Delta_1^2\neq 0$ $\Delta_2^2>\Delta_3^2$.
In a vicinity of the singular point $\tau_A=\tau_B$ of particular interest are combinations of the coefficients $\beta$, which tend to zero in the weak coupling limit. The expression for the splitting $\tau_A-\tau_B$ in the leading order depends on two such combinations $\beta_{45}\equiv\varepsilon\beta_2$ and $2\beta_1+\beta_3\equiv\nu\beta_2$:   $(\tau_A-\tau_B)\approx\frac{9}{2}\kappa(\varepsilon-\nu)$.
Combination of Eqns. (11) and (12) renders
$$
[\beta_{34}-(2\beta_1+\beta_{5})]\Delta_3^2=[(\nu-\varepsilon)\Delta_0^2+(\nu+\varepsilon)(\Delta_2^2-\Delta_1^2)]\beta_2.        \eqno(15)
$$
At finite $\varepsilon$ and $\nu$ $\Delta_3^2\neq 0$ so that the new phase is strictly speaking non-ESP and not axi-planar.
In a principal order on $\varepsilon\ll 1$ and $\nu\ll 1$:
$$
\Delta_3^2=-\frac{1}{16\beta_2}\left[(3\nu-\varepsilon)\tau+(9\nu-19\varepsilon)\frac{\kappa}{2}\right].        \eqno(16)
$$
Presence of the term, proportional to $\tau$ in this expression means that the order parameter of the new phase remains distorted even in the limit $|\tau|\gg|\kappa|$. The distortion is relatively small only because $\varepsilon$ and $\nu$ are small.

\textbf{5.} Now we compare the expected NMR properties of the new phase with the properties of the ESP2-phase of Ref. \cite{askh}. In these experiments different phases are characterized by their transverse NMR frequency shifts as functions of the tipping angle $\beta$ and the angle $\mu$ between the direction of the d.c. magnetic field and the anisotropy axis $z$  The calculated dependence of the shift on these angles for the order parameter (8) with $\cos\theta=0$ in the two-dimensional  Larkin-Imry-Ma state is given by the formula:
$$
\chi\omega_L\Delta\omega=const.\left(2|\Delta_0|^2-|\Delta_1|^2\right)\left[\cos\beta-\frac{\sin^2\mu}{4}(5\cos\beta-1)\right]+
$$
$$
|\Delta_2|^2\left[\frac{\sin^2\mu}{4}(7\cos\beta+1)-\cos\beta\right]+
|\Delta_3|^2\cos\beta\left[2-3\sin^2\mu\right].       \eqno(17)
$$
Here the proportionality coefficient is the same for all phases and it is determined by the properties of the normal phase, $\omega_L$ is the Larmor frequency and $\chi$ - the maximum principal value of a tensor of the magnetic susceptibility of the considered phase.
As a zero order approximation in $\varepsilon$ and $\nu$ we use the order parameter $A_{\mu j}^{AP}$ with temperature dependencies of $\Delta_0$, $\Delta_1$, $\Delta_2$ as given in the paragraph, following Eq. (9).  Then
$$
\chi\omega_L\Delta\omega_{AP}\sim-\frac{1}{2}\left(\tau+\frac{19}{2}\kappa\right)\cos\beta+
\frac{\sin^2\mu}{16}\left[(7\cos\beta-5)\tau+\frac{\kappa}{2}(205\cos\beta-23)\right].             \eqno(18)
$$
For comparison, in the distorted ABM-phase (ESP1-phase of Ref.\cite{askh}):
$$
\chi\omega_L\Delta\omega_{ABM}\sim-\frac{1}{2}\left(\tau+\frac{19}{2}\kappa\right)
\left[\cos\beta-\frac{1}{4}(\sin^2\mu)(5\cos\beta-1)\right].             \eqno(19)
$$
When magnetic field is parallel to $z$ ($\mu$=0) the calculated transverse NMR shift for the axi-planar phase coincides with that for the ABM-phase. In the experiments \cite{askh} the shifts for ESP1 and ESP2 phases do not coincide, but the difference is not big and it can be ascribed to the corrections of the order of $\varepsilon$ and $\nu$.  More definitive experimental check of the proposed identification would be a measurement of the NMR shifts at the  perpendicular orientation of the field to the anisotropy axis ($\sin^2\mu$=1). For this orientation two shifts have to be different -- in the distorted ABM-phase CW shift is zero, while in the axi-planar phase
$$
\chi\omega_L\Delta\omega_{AP}\sim-\frac{3}{8}\left(\tau-\frac{5}{2}\kappa\right).    \eqno(20)
$$
The ratio of the shift for $\sin^2\mu$=1 to that for $\sin^2\mu$=0 in this phase has to be  $\frac{3(2\tau-5\kappa)}{4(2\tau+19\kappa)}$ at $(2\tau-5\kappa)<0$. The data for the ESP2 phase at $\sin^2\mu$=1 are not yet available.

In conclusion, we suggest to identify the ESP2 phase, observed in the experiments \cite{askh} as the phase with the order parameter Eq. (8) at $\cos\theta=0$. In the principal order on small deviations of coefficients $\beta_s$ from their weak coupling values it reduces to the axi-planar form $A_{\mu j}^{AP}$ given by Eq. (9). The proposed identification can be checked in NMR experiments.
The performed analysis of possible phase transitions in the Ginzburg and Landau region can be used also for extraction of values of coefficients $\beta_s$ from the experimental data with a subsequent quantitative interpretation of the data. A more extensive experimental investigation in particular in the low pressure region would be useful for realization of this program.

We thank V. V. Dmitriev for stimulating discussions and useful comments. This research was supported in part by the Russian Foundation for Basic Research under project  11-02-00357-a.

\end{document}